\documentclass[twocolumn,10pt]{asme2ej}

\usepackage{graphicx} 
\usepackage{hyperref} 
\usepackage[normalem]{ulem}  

\hypersetup{
	colorlinks=true,
	linkcolor=blue,
	citecolor=blue,
	urlcolor=blue,
}
\usepackage[square,numbers]{natbib}
\usepackage{multirow}
\usepackage{amssymb,amsmath}
\usepackage{graphics}
\usepackage{enumitem}
\usepackage{bm,comment}
\usepackage{lineno}
\usepackage{xcolor}
\usepackage{caption} 
\usepackage{array}     
\usepackage{float}     
\title{Towards Smart Manufacturing Metaverse via Digital Twinning in Extended Reality}

\author{Hui Yang\thanks{Corresponding author. Email: huy25@psu.edu.} 
    \affiliation{
        Industrial and Manufacturing Engineering\\
	The Pennsylvania State University\\
	University Park, Pennsylvania\\
        Email: huy25@psu.edu
    }
}

\author{Faisal Aqlan
    \affiliation{
	Industrial and Systems Engineering\\
    University of Louisville\\
    Louisville KY, USA\\
    Email: faisal.aqlan@louisville.edu
    }
}

\author{Richard Zhao
    \affiliation{
	Computer Science\\
    University of Calgary\\
    Calgary AB, Canada\\
    Email: richard.zhao1@ucalgary.ca
    }
}

\begin{document}
\maketitle    

\begin{abstract}
{\it The rapid evolution of modern manufacturing systems is driven by the integration of emerging metaverse technologies such as artificial intelligence (AI), digital twin (DT) with different forms of extended reality (XR) like virtual reality (VR), augmented reality (AR), and mixed reality (MR). These advances confront manufacturing workers with complex and evolving environments that demand digital literacy for problem solving in the future workplace. However, manufacturing industry faces a critical shortage of skilled workforce with digital literacy in the world. Further, global pandemic has significantly changed how people work and collaborate digitally and remotely. There is an urgent need to rethink digital platformization and leverage emerging technologies to propel industrial evolution toward human-centered manufacturing metaverse (MfgVerse). This paper presents a forward-looking perspective on the development of smart MfgVerse, highlighting current efforts in learning factory, cognitive digital twinning, and the new sharing economy of manufacturing-as-a-service (MaaS). MfgVerse is converging into multiplex networks, including a social network of human stakeholders, an interconnected network of manufacturing things or agents (e.g., machines, robots, facilities, material handling systems), a network of digital twins of physical things, as well as auxiliary networks of sales, supply chain, logistics, and remanufacturing systems. We also showcase the design and development of a learning factory for workforce training in extended reality. Finally, future directions, challenges, and opportunities are discussed for human-centered manufacturing metaverse. We hope this work helps stimulate more comprehensive studies and in-depth research efforts to advance MfgVerse technologies.\\}
\end{abstract}



\section{Introduction} \label{sec1:intro}
The evolution of manufacturing systems has been driven by the need for increased efficiency, flexibility, precision, and sustainability. As the industrial revolution progresses, the integration of cutting-edge technologies such as new human-machine interfaces (HMIs) and metaverse is becoming vital to transform traditional manufacturing practices. As opposed to traditional 2D screens and monitors, emerging HMIs enable interactions in the 3D immersive virtual environment through different forms of eXtended reality (XR), including virtual reality (VR), augmented reality (AR), and mixed reality (MR). \textbf{Metaverse }results as \textit{the convergence of physical reality and virtual world, providing a persistent, immersive, scalable, and shared digital space (or environment) for people to interact and work with physical and/or virtual beings}. Current metaverse developments include virtual gaming worlds (e.g., Roblox, Fortnite), social spaces (e.g., Meta’s Horizon), or work platforms (e.g., Microsoft Mesh). For instance, Roblox has 88.9 million active daily users in 2025 \cite{roblox2025}. The global metaverse market is estimated to exceed 1.1 trillion by 2030, with 80\% of users under 16 \cite{metaverse2025b}. Even if some users are offline, metaverse continues to exist and evolve in the cyberspace. More than 60\% of users spend time on non-gaming activities, e.g., working, collaborating, and/or performing transactions in the online marketplace \cite{metaverse2025a}. This shows a strong promise for developing a new manufacturing metaverse (MfgVerse) for the next generation of digital natives.  

In fact, rapid technological advances confront manufacturing workers with complex and evolving environments that demand digital literacy for problem solving in the workplace \cite{fischer2012process}. Nearly 1.4 million American jobs will be disrupted by technological advances by 2026 \cite{world2018towards}. In the next decade, U.S. manufacturing alone will not have enough skilled workforce to fill 2.4 million jobs \cite{giffi20182018}. Globally, technological advances will drive 14\% of the workforce to switch job categories away from manufacturing \cite{ellingrud2020building}. Moreover, global pandemic like 2019 Coronavirus (COVID-19) has changed almost every aspect of our life, including how people work, collaborate, and perform manufacturing operations and workforce training \cite{yang2021epidemic,liu2021spatialtessell}. This creates an urgent need to rethink digital platformization and develop new metaverse platforms that utilize emerging technologies to support human-centered manufacturing \cite{10.1115/1.4068082}. This represents a pressing challenge to stakeholders in academia and industry who need to prepare future-generation workforce in manufacturing.

By creating an immersive and interactive environment between physical and digital worlds, metaverse technologies are bringing significant transformation to the industry toward human-centered digital manufacturing. This, in turn, enables immersive simulations, real-time monitoring, process control, and dynamic adaptation to operator needs, which promotes remote collaboration and productivity among geographically distributed teams. By bridging the gap between physical operations and digital twins (DTs), MfgVerse represents the next revolution of manufacturing industry beyond Industry 5.0 \cite{yang2019internet}. This presents an unprecedented opportunity to disrupt manufacturing practice in ways that were previously unimaginable. For example, advanced sensing, control systems, and HMIs enhance MfgVerse by dynamically adjusting to user behavior and environmental factors for effective interaction between humans and machines through optimal adjustments to user needs and situational demands.

	\begin{figure*}[htb]
		\centering
		\includegraphics[width=6.5 in]{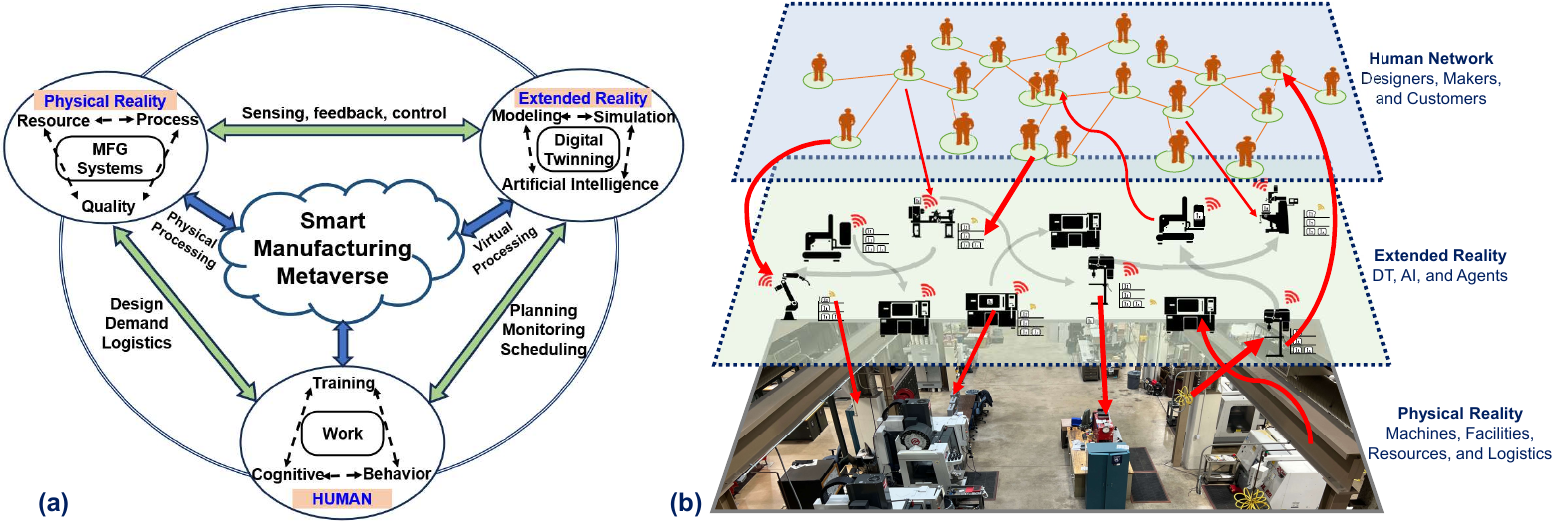}
        \caption{Illustrations of (a) the interconnected manufacturing metaverse and (b) multiplex relationships among human network, extended reality and physical reality.}
		\label{fig:socialmfg}
	\end{figure*}

In contrast with conventional manufacturing supply chains, digital platformization can connect customers needing personalized products with manufacturing service providers via a two-sided market in the metaverse environment. Note that traditional supply chains are often linear and lengthy, involving a multi-stage process from the design and prototyping to manufacturing and distribution to warehousing and retailing, thereby reaching end users or customers. However, MfgVerse offers an unprecedented opportunity to fully realize the sharing economy of manufacturing as an on-demand service (MaaS) \cite{yang2021stable}. As a result, service providers can increase the utilization rate of their idle assets (e.g., materials, machines, 3D printers, microfactories, even maintenance and support). This will greatly circumvent the lengthy product lifecycle and also promote the circular economy through the resell, reuse, and recycling of second-hand products in the online marketplace. 

For instance, a client will no longer need to search and find custom parts in traditional brick-and-mortar stores \cite{yang2021stable}. Instead, customers can create and optimize the design by learning and working with others through the metaverse. Once finalized, this design can be transferred online and matched with an idle asset (e.g., a 3D printer) for on-demand production. Through XR and DT technologies, human stakeholders pertinent to this product can visualize, monitor, and control the manufacturing process. After manufacturing and inspection, the client will receive custom parts and pertinent documentations or reports in the mailbox. In this way, MfgVerse propels the Industry 5.0 evolution and brings significant benefits as follows:

\begin{itemize}
    \item [-] \textbf{Remote collaboration}: 3D immersive environments for virtual meetings with spatial awareness;
    \item [-] \textbf{Design \& prototyping}: Remote work and prototyping among geographically dispersed teams via the overlay of digital designs on physical parts;
    \item [-] \textbf{Manufacturing operation}:	DTs for dashboard visualization, process monitoring, quality control, networked coordination, and job scheduling;
    \item [-] \textbf{Maintenance \& support}: Remote AR to display step-by-step procedures on top of real machines for repair and troubleshooting;    
    \item [-] \textbf{E-commerce \& logistics}: Virtual walkthrough the value chain, online transactions, product tracking, and supply chain optimization;
    \item [-] \textbf{Virtual learning factory}: A safe, persistent, and 3D immersive learning environment for manufacturing training and education.
\end{itemize}

The objective of this paper is to present a forward-looking perspectives on sensor-based innovations in XR, DT, AI toward a new paradigm of smart manufacturing metaverse. The emergence of new HMIs changes the way people interact with machines, internet, and computing devices from traditional 2D screens to the 3D immersive cyber space with physical senses. As shown in Fig. \ref{fig:socialmfg}, \textit{manufacturing metaverse is converging into multiplex networks, including a social network of human stakeholders, an interconnected network of manufacturing things (e.g., machines, facilities, material handling systems), a network of DTs of physical things, as well as auxiliary networks of sales, supply chain, logistics and remanufacturing systems to name a few}. In cyberspace, DTs are boosted with rapid advances in AI, simulation optimization, and big data analytics to improve the smartness level of manufacturing operations.  We present a new framework of cognitive digital twin (CDT) to embed agentic AI capabilities in DT models of physical twins, things, or agents. We also showcase the design and development of virtual learning factory for workforce training, Finally, we identify critical challenges and propose a research agenda to drive the evolution of human-centered manufacturing metaverse.  We hope that this work will fuel increasing interests in multidisciplinary research, as well as the novel design and development of new manufacturing metaverse technologies.

The remainder of this paper is organized as follows: Section \ref{sec:socialmm} discusses the context and definition of manufacturing metaverse, and discusses technologies and principles that drive its evolution, including human-machine interfaces (HMIs) and cyber-physical twinning. Section \ref{sec:m1m4} delves into the VR factory - simulating the manufacturing evolution (e.g., from craft production to mass production, mass customization, and personalized manufacturing) in the virtual environment. Section \ref{sec:CDT} presents the integration of DT and AI technologies, including physical and digital twins, embedded agentic AI, human-centered CDT, and generative intelligence. Section \ref{sec:future} highlights future directions, challenges, and research opportunities. Finally, Section 9 concludes with a forward-looking summary of insights and hopes to catalyze more research and development efforts in the community.

\section{Smart and Interconnected Manufacturing Metaverse} \label{sec:socialmm}
New generations of manufacturing workforce grow up in an era of metaverse, XR, smartphones, social media, and instant access to information. As a result, they are often considered “\textit{digital natives}”, who value individuality and expect new learning experiences tailored to their needs. There is an urgent need to understand and accommodate the preferences and priorities of new generations in the evolving landscape of digital manufacturing. Fig. \ref{fig:socialmfg} shows \textbf{manufacturing metaverse} \textit{connects a network of human users, e.g., designers, makers, and consumers via the collaborative virtual environment (e.g., XR factory), and further operates the DTs of heterogeneous agents (e.g., AIs, machines, material handling systems, facilities, resources) for the provision of smart manufacturing as an on-demand service}.

This new MfgVerse framework is inspired by our previous investigation of sharing economy in a digital fabrication network \cite{yang2021stable}. We studied the stable matching between human designers or makers and a network of underutilized 3D printers across the 20 commonwealth campuses in a University setting \cite{yang2021stable}. The new MaaS business model prioritizes utilization over ownership within a two-sided marketplace, where manufacturing resources are shared on-demand as a service through a digital platform. As a vertical step, MfgVerse is embodied with three core components, physical manufacturing systems, DTs in the VR environment, and human stakeholders (e.g., operators, engineers, designers, technicians) in an integrated framework. Key benefits of MfgVerse include:
\begin{itemize}
    \item [-] \textbf{Physical manufacturing systems}: In the physical world, there are often underutilized resources such as machines, production lines, microfactories, sensors, robots, warehouses. MfgVerse allows users to remotely access a large network of manufacturing resources for production needs without owning or operating the machine shops or factories. 
    \item [-] \textbf{Digital twinning in the virtual environment}: DT is a digital representation of manufacturing operations in the cyberspace of metaverse through modeling and simulation (i.e., enabling the dashboard visualization, statistical process monitoring \cite{krall2023virtual}, job tracking), while analytics in cyberspace distill actionable insights and optimize decision making (e.g., manufacturing planning, job scheduling, process control) in the physical twin.
    \item [-] \textbf{Human-centered manufacturing in the meataverse}: Human stakeholders can interact with each other and manufacturing resources via a collaborative, immersive, and e-commerce metaverse that combines XR, AI, DT and the internet. This also offers a virtual learning factory for the training and education of next-generation manufacturing workforce.
\end{itemize}


\begin{table*}[t]
    \centering
        \begin{tabular}{m{6.0in}}
            \includegraphics[width=6.0in]{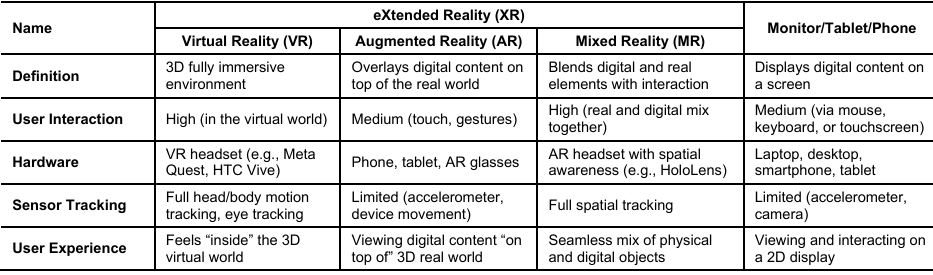} \\
        \end{tabular}
    \caption{Comparison among different HMIs -- VR, AR, XR, and monitor/tablet/phone.}
    \label{tab:hmicompare}
\end{table*}

In a nutshell, we envision a new form of sharing economy via MaaS and social networks that creates new opportunities for entrepreneurship. With affordable and user-friendly manufacturing technology offered as on-demand services, individuals and communities are enabled to enter niche markets and bring innovative ideas to life. The MaaS network connects local fabrication hubs with distributed machines, paving the way to turn under-utilized assets into productive resources. In addition, social metaverse facilitates the planning and control of manufacturing activities in the networked virtual environment. This, in turn, helps streamline production processes and optimize resource utilization. As a result, smart MfgVerse impacts beyond individual businesses to encompass broader economic growth, job creation, and innovation across sectors.

\subsection{Human-machine interface (HMI)} \label{sec:HMI}
Emerging VR, AR, and MR technologies provide new ways for human to interact with machines and computers in the manufacturing environment. As opposed to 2D screens or monitors, immersive 3D HMIs (e.g., Meta's Quest, Apple’s Vision Pro, Microsoft Hololens, smart glasses) will dominate remote work and collaboration for future manufacturing and design. Although legacy monitors, tablets or phones remain useful and practical tools for HMI tasks, they are non-immersive 2D interfaces that are dependent on keyboard, mouse, or touch as the medium for interaction. As shown in Table \ref{tab:hmicompare}, there are key differences among VR, AR, XR, and conventional 2D screens or monitors. 
\begin{itemize}
    \item [-] \textit{VR} engages the users inside the 3D immersive virtual environment, which often consists of avatars, assets and resources built as digital replicas of physical beings in the real world. Users wear a headset or similar device (e.g., Quest, HTC Vive) that allows them to interact with and explore this virtual environment. VR factory is best suited for workforce training and manufacturing simulations, but requires sensors such as accelerometers, motion or eye tracking for cyber-physical interactions. 
    
    \item [-] \textit{AR} overlays digital content onto the view of real-world assets via a smartphone, tablet, or AR glasses. AR does not block out the physical world as VR, but rather enables human operators view digital contents around physical assets. For e.g., maintenance engineers can use AR glasses to display the repair procedures step-by-step "on top of" a failed machine. AR augments the user’s perception of reality by adding digital contents, such as graphics or text, to the real-world environment.

    \item [-] \textit{MR} is an upgraded version of AR which further allows real-time interactions between digital and physical objects or things. For example, manufacturing engineers can leverage advanced headsets like Microsoft HoloLens to overlay 3D holographic models of machine components (e.g., shaft, engine, cutting tools) onto physical machines, allowing the manipulation of digital objects to interact with the real world. MR devices can track the user’s movements and adjust the positions, rotations, and lighting of 3D virtual objects accordingly, creating a more immersive experience. MR can be utilized to simulate environments and provide visual and auditory stimuli to assess individuals’ response to different scenarios in learning factories for process improvement \cite{juraschek2018mixed}.    

\end{itemize}

	\begin{figure*}[hbt]
		\centering
		\includegraphics[width=5.0 in]{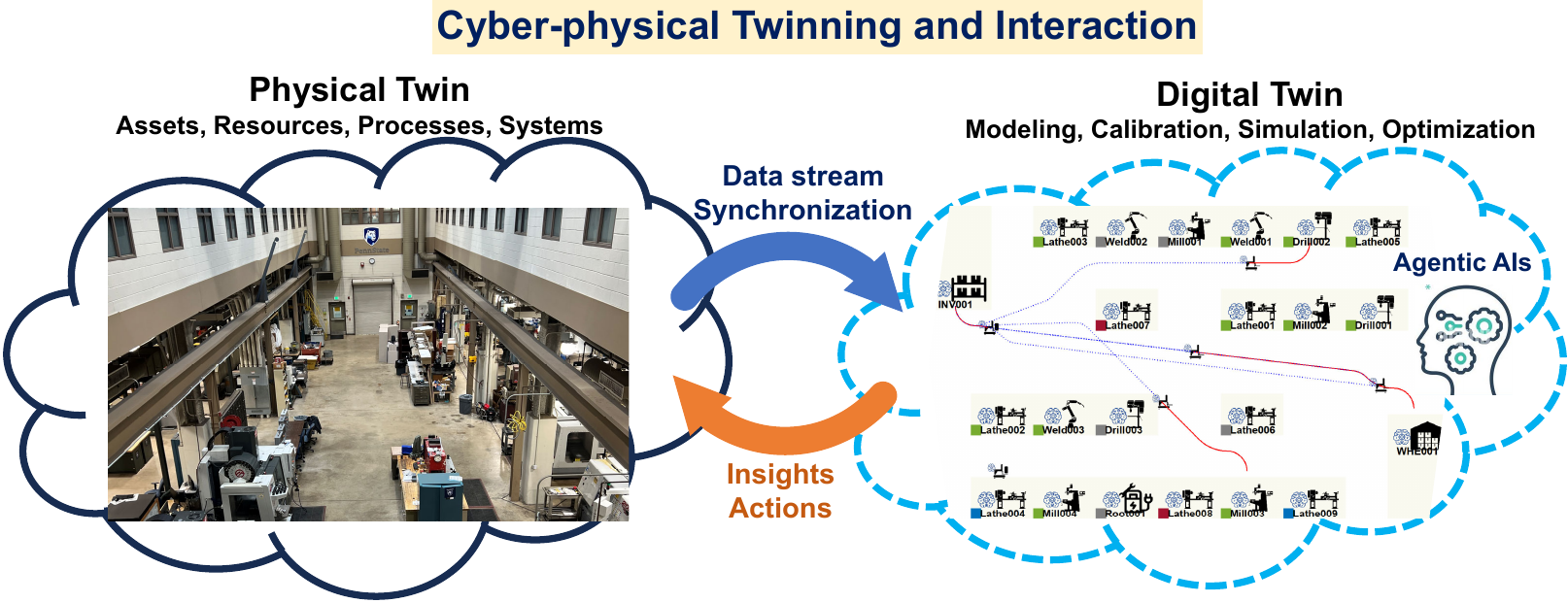}
		\caption{Illustration of cyber-physical twinning and interaction embodied by three core components -- physical twin in the real world, DT in virtual space, and interactions between physical and digital twins.}
		\label{fig:cpti}
	\end{figure*}

Further, XR describes the immersive technologies encompassing real-and-virtual combined environments as well as human-machine interactions generated by sensing technology and wearables. In general, XR is an umbrella term that covers VR, AR, and MR. Immersive HMI technologies bring the disruption to future manufacturing, posing a fundamental shift in the way people will interact with technology and media. Hence, a new set of professional skills will be required. XR simulations are becoming an essential part of manufacturing workforce training. There is significant evidence that VR can help students learn faster and better retain information \cite{di2013impact}.  XR can improve learning outcomes by providing a richer context and allowing more realistic situations than what students can experience in a classroom setting. A study conducted an in-depth analysis on whether people learn better through virtual, immersive environments as opposed to more traditional platforms. The study found that about 9\% improvement in information recall accuracy can be achieved when using VR headsets \cite{krokos2019virtual}. AR combines the elements from both physical and virtual environments. AR simulations enhance the user’s sensory perception of the real world with a contextual layer of information \cite{azuma1997survey}. Therefore, physical, VR, AR, MR simulations will guide the development of cyber-physical XR simulations. Sensing technologies that include physiological sensors, eye and motion tracking will further provide real-time monitoring of human behavior and dynamics, thereby contributing to the digital twinning process.

\subsection{Cyber-physical Twinning} \label{sec:CPT}
DT involves a virtual representation of physical manufacturing assets, resources, processes, or systems that enables real-time data streaming, dashboard visualization, predictive analytics, and intelligence distillation \cite{lee2023digitala,lee2023digitalb}. Broadly, there are three core components in the DT -- physical twin in the real world, DT in virtual space, and interactions between physical and digital twins. The twinning and modeling process is non-trivial, often encountering data and model challenges such as sensor placement, missing data, model uncertainty \cite{tao2019make,tao2022digital}. XR, including VR, AR, and MR, has gained popularity in recent years as a new tool to support DT modeling in 3D immersive environments. The integration of XR and DT can also provide a safe and controlled environment for manufacturing simulations in workforce training. 

Here, we argue that \textit{``digital twin" means a twin in the cyberspace, while ``physical twin" means a twin in the physical world. The term of "digital twin" does not explicitly cover three core components -- physical twin, digital twin, and interactions between physical and virtual twins}. In fact, cyber-physical interaction is indispensable to derive data-driven intelligence and simultaneously feed actions or insights back to the physical twin of a real-world factory. Therefore, \textbf{cyber-physical twin} (or Phygital Twin) is a more appropriate term that refers to \textit{the process to build the digital twin with data from a physical twin, and further enable cyber-physical interactions through data synchronization, online calibration, dynamic learning, and artificial intelligence}. 

As shown in Fig. \ref{fig:cpti}, advanced sensing systems (e.g., supervisory control and data acquisition - SCADA) are critical to streaming the data or signals from the physical twin to the digital twin. The quality of data is instrumental to modeling, calibration, simulation, and optimization in the digital twin. In turn, the integration of sensors with DT offers an unprecedented opportunity of 3D immersive environment to visualize the operational statuses of machines and engage in real-time interactions between digital and physical machines. Further, analytics in the DT help optimize managerial decisions such as job scheduling, process control, predictive maintenance, and feed actions back into the physical twin. After all, smart manufacturing operations depend greatly on the seamless cyber-physical twinning and integration.

 \subsection{Sensor Integration for Smart Metaverse} \label{sec:sensor}
Indeed, advanced sensing is indispensable to effective cyber-physical twinning and interaction. MfgVerse depends on the design of advanced sensor technologies (e.g., eye tracking, motion tracking, and voice tracking) to improve immersive experience, implementation efficiency, and user engagement. For example, our prior work integrated an HTC Vive headset with a custom-installed Tobii eye tracker and logged 24 participants who trained in a virtual factory \cite{zhu2022sensor}. Then, we developed a Signal-Detection-Theory model based on eye-tracking data to study problem-solving behaviors in the virtual environment. While an AI assistant delivers context-sensitive guidance, our experiments showed how sensor signals help close the perception–action loop and provide real-time personalization of instruction timing and verbosity \cite{sloan2022adaptive}. Fu et al. created an ergonomic assessment pipeline for a human-robot collaboration workstation in modular construction, which leverages hand tracking for interactive manipulation in Unity and two video cameras for full-body kinematics \cite{fu2025virtual}. Shishir and Zhao evaluated the Meta Quest Pro’s onboard eye tracker in a gaze interaction that allow users to select distant objects using the Jedi-Force-Like Pull method with a pinch gesture \cite{shishir2025jedi}.  Aslam and Zhao built a voice-augmented interface in Unity, and utilized the Wit.ai speech recognition service to parse each utterance into an intent-entity-trait triple (e.g., “rotate right at thirty degrees") \cite{aslam2024voice}. Seven canonical manipulation tasks (i.e., scaling, locomotion, selection, zooming, rotation, positioning, and collapsing a 3D model) were benchmarked between speech interaction and controller interaction methods for 26 participants. The speech method outperformed controller interactions in every task with statistical significance at 95\% confidence level, and reduced error and mistake counts by an order of magnitude.

	\begin{figure*}[hbt]
		\centering
		\includegraphics[width=6.0 in]{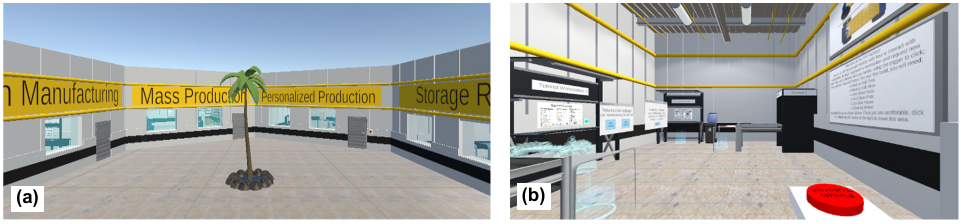}
		\caption{Example snapshots of the virtual factory: (a) the factory hub, and (b) the tutorial room.}
		\label{fig:vrfactory1}
	\end{figure*}
    
Further, Salinas-Martínez et al. implemented a multimodal human-robot interaction platform \cite{salinas2024multimodal} for printed-circuit-board manufacturing that fuses three sensor channels, including (i) \textit{Leap Motion Controller}: infrared cameras and sensors capable of highly accurate tracking of hand movements; (ii) \textit{Tap Strap 2}: a wearable that detects forearm micromovements for subtle pinching; and (iii) \textit{Cellya speech interface}: a microphone array using Mozilla DeepSpeech for natural language commands. Peruzzini et al. developed a multi-sensor virtual manufacturing cockpit that integrated an HTC Vive Pro Eye HMD for viewpoint tracking, an Xsens inertial suit for full-body capture, and a Leap Motion infrared sensor for gesture recognition \cite{peruzzini2021using}. The hybrid rig captured gross postures using Xsens and fine hand gestures through Leap Motion, and the Vive Pro Eye provided on-device eye tracking, enabling real-time analysis of assembly ergonomics and human-machine interaction times. 

Multimodal sensing turns XR HMIs from immersive visualization tools into a broad set of measurement instruments of physical, cognitive, and affective operator states. As such, XR HMIs use real-time sensor data to learn and improve user experiences on the fly, enhancing usability, efficiency, and comfort. Integrating sensors into XR systems helps respond to diverse inputs and continually personalize interactions and workflows, for example,

•  \textit{Rule-based ergonomic adaptation}: Predefined thresholds on joint angles or gaze dwell time trigger interface changes such as repositioning virtual instructions or slowing a collaborative robot movement trajectory. Khamaisi et al. showcased these rules in a human-cyber-physical-system design framework that iteratively updates the DT of the workstation \cite{khamaisi2025designing}. The role of sensor-driven HMIs is emphasized when improving industrial tasks through real-time feedback to human inputs.

• \textit{Physiology-aware visual modulation}: Closed-loop engines alter environmental parameters (background clutter, brightness, color saturation, or information density) according to arousal proxies such as GSR (Galvanic Skin Response) and HRV (Heart Rate Variability). Chiossi et al. showed that increasing visual complexity only when sensors indicated low arousal preserved user performance while preventing overload \cite{chiossi2023senscon}.

• \textit{Sensor-driven prediction and co-adaptation}: Sensor-based models help forecast upcoming workload or intent and preemptively adjust the interface. Examples include LSTM networks that predict the grasp force from EMG to select tool parameters, and transformer-based speech recognition models that anticipate ambiguities in spoken commands \cite{salinas2024multimodal}. Early field evaluations report cycle-time reductions and measurable drops in NASA-TLX workload scores, although most studies are still confined to laboratory pilots. See a comprehensive review of time series forecasting methods in \cite{yang2011local, cheng2015time}.

\section{Simulating the Manufacturing Evolution in the Virtual Environment} \label{sec:m1m4}

Our team has designed a virtual factory to simulate and explore the evolution of manufacturing paradigms in an immersive environment \cite{zhu2022sensor,aqlan2021virtual,kim2024behavioral}. The virtual factory serves as a dynamic educational and research tool, which allows users to engage with historical and contemporary manufacturing systems. By virtually experiencing the transition from manual craftsmanship to data-driven personalization, participants gain a deeper understanding of how innovation, efficiency, and customization have shaped manufacturing practices over time. Sample snapshots from the virtual factory are shown in Fig. \ref{fig:vrfactory1}. The multi-user setting allows multiple individuals to use the virtual platform simultaneously, and they can be in different geographical locations.

    \begin{figure*}[hbt]
		\centering
		\includegraphics[width=5.5 in]{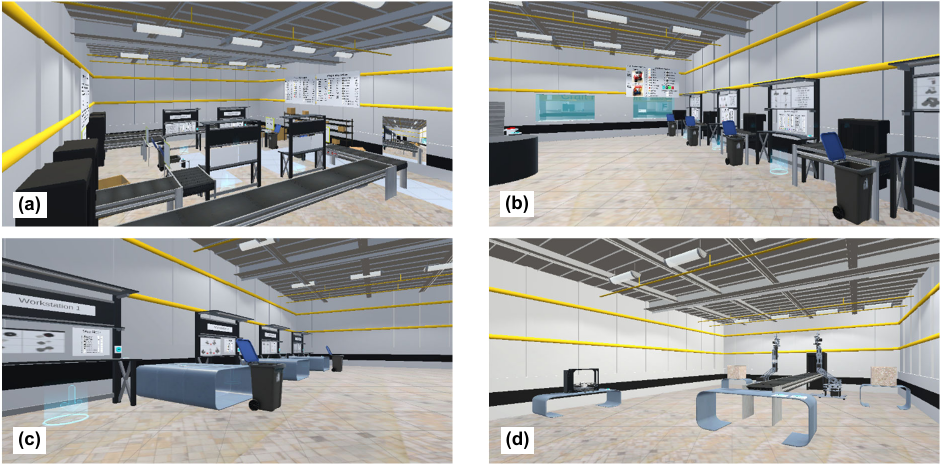}
		\caption{Example snapshots from four rooms in the virtual factory that represent the manufacturing paradigms: (a) craft production, (b) mass production, (c) mass customization, and (d) personalized production. See the AR demo of users assembling cars while overlaying digital contents on top of the real world in this YouTube video: \url{https://youtu.be/EApLtKmnf2U}.}
		\label{fig:vrfactory2}
	\end{figure*}

The manufacturing industry has continuously evolved through distinct paradigms - ranging from craft production to mass production, mass customization, and now personalized manufacturing \cite{koren2010global}. Each evolution has been marked by transformative changes in technology, processes, and workforce roles. Understanding these shifts is critical for preparing future workforce to navigate modern manufacturing landscapes. Industry 1.0 (1850-1913) involves the move from home-based crafting activities to production in factories with centralized energy sources that powered manufacturing machinery \cite{sabel1985historical}. The term “craft production” is associated with this revolution. Every unique product is produced separately in a small machine shop supported by limited technology aids and general-purpose machines. High-quality products are produced, usually at high costs and without any standardization \cite{modrak2014handbook}. 

Industry 2.0 (1913-1980) progresses from individual piecework to mass production. The development of integrated systems, such as assembly lines and automation, expedited high-volume production of identical parts cost reduction \cite{yang2013continuous}. Industry 3.0 (1980-2010) is driven by the introduction of numerical machine control (CNC), later computerized control, and process automation, providing increased accuracy and flexibility and facilitating the execution of lower-volume-higher-mix manufacturing scenarios. Customized products are produced with near mass production efficiency \cite{tseng2001mass}. Modern manufacturing is in the midst of the fourth revolution that integrates networked cyber-physical systems to analyze and optimize production processes. Personalization is an emerging paradigm associated with this revolution, whereby customers can tailor products to their individual needs while maintaining high production efficiency. Product options are designed by customers, sold and produced using advanced manufacturing and service systems \cite{koren2010global}.

	\begin{figure*}[hbt]
		\centering
		\includegraphics[width=5.5 in]{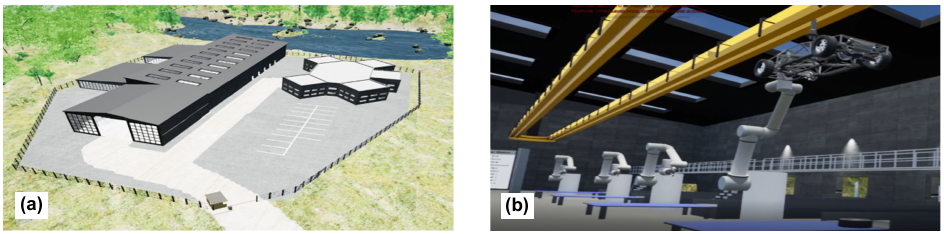}
		\caption{Example snapshots from the mass customization factory: (a) a bird-eye view of the factory, and (b) a view from the automated assembly room showing robots assembling car components (also see the VR demo of assembly operations in this YouTube video: \url{https://youtu.be/i5nVBOFaeWw}).}
		\label{fig:vrfactory3}
	\end{figure*}

	\begin{figure*}[hbt]
		\centering
		\includegraphics[width=6.5 in]{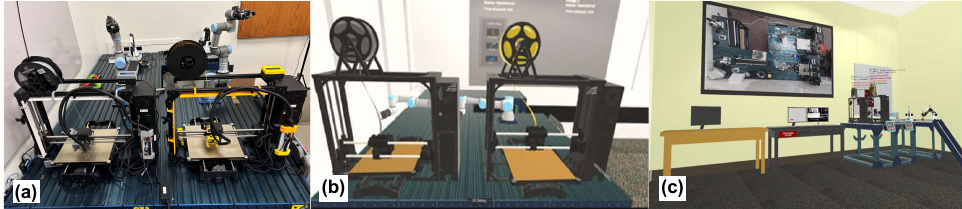}
		\caption{Example snapshots of robotic arm and 3D printer systems: (a) physical twin, (b) digital twin, (c) real-time live camera feed of physical operations in the virtual environment. See the DT-integrated MR demo of real-time interactions between cyber-physical twins in \url{https://youtu.be/O8vevgfxxwQ}.}
		\label{fig:exampleDT}
	\end{figure*}

As shown in Fig. \ref{fig:vrfactory1}, the virtual factory takes the production of car toys as a learning case study of automotive manufacturing \cite{zhu2022sensor, aqlan2021virtual, kim2024behavioral}. The virtual environment implements team-based performance requirements that must be met for the group to advance through the different manufacturing paradigms. This structure encourages collective responsibility and reinforces the importance of achieving shared goals. The virtual factory is also equipped with advanced sensing technologies such as eye-tracking \cite{zhu2022sensor} and facial expression analysis tools. These tools capture detailed behavioral and physiological data during collaboration, which enable researchers and instructors to analyze team dynamics, communication efficiency, attention levels, and emotional responses \cite{kim2024behavioral}.

Fig. \ref{fig:vrfactory2} shows sample snapshots from the rooms in the virtual factory that represent the variety of manufacturing paradigms. The first room (Fig. \ref{fig:vrfactory2}a) represents the craft production environment and features four independent assembly stations. Each station is equipped with a dedicated set of tools and virtual components, allowing participants to individually work on the customization of product builds.  The second room (Fig. \ref{fig:vrfactory2}b) represents the mass production environment, which consists of four sequential workstations connected by conveyor belts that facilitate the movement of parts from one station to the next. The third room (Fig. \ref{fig:vrfactory2}c) represents the mass customization paradigm, which combines the efficiency of mass production with the flexibility of custom design. It allows customers to choose from various options for car parts such as wheels, roofs, and windshields, while maintaining high production volume and low costs. The personalized production room (Fig. \ref{fig:vrfactory2}d) enables users to co-design products tailored to their individual needs. Unlike mass customization, which limits choices to predefined options, personalized production allows active customer involvement in the design process.

We further developed specific virtual environments to simulate mass customization and personalized production. The mass customization factory, shown in Fig. \ref{fig:vrfactory3}, allows users to produce a variety of car models that combine personalized features with standardized components \cite{aqlan2021virtual}. Assembly operations are performed through both manual and automated processes, which simulate real-world hybrid production lines (also see the YouTube video: \url{https://youtu.be/i5nVBOFaeWw}). The manufacturing processes include welding, fastening, 3D printing, machine painting, supply and demand, and six-sigma defect checking \cite{yang2020six}. The personalized production environment includes a VR-based DT of a physically automated station that consists of 3D printers and robots (see Fig. \ref{fig:exampleDT}). The DT enables users to participate in end-to-end product creation, from initial design decisions to real-time fabrication. Users interact with the system to set print coordinates, initiate prints via the user interface in the VR environment, and monitor process variables such as temperature, humidity, and print quality \cite{yang2023multiresolution}. Physical printers execute these commands in parallel to produce parts that are then retrieved and assembled by robots. These robots autonomously transfer the assembled components to a collection area using conveyor mechanisms. 

In particular, it is worth noting that this virtual factory extends beyond VR to include AR and MR components. The digital contents are overlayed on top of the real world via the Microsoft HoloLens. The physical lab and room are shown in the background of the virtual factory, where the users are experiencing the immersive virtual environment in the presence of physical world. See the AR demo of users assembling cars while overlaying digital contents on top of the real world in this YouTube video: \url{https://youtu.be/EApLtKmnf2U}. Also, MR enables real-time interactions while blending the digital and real elements in the virtual factory. Advanced sensors are integrated with immersive 3D environment to enable cyber-physical interactions (also see details in Section \ref{sec:CDT}: CDT). As such, DT connects collaborative robots and 3D printers, allowing users to tele-operate machines in the metaverse. Real-time sensor data are synchronized for analytics, while robots assemble printed parts into final products. See the DT-integrated MR demo of users operating robots and 3D printers with real-time interactions between cyber-physical twins in this YouTube video: \url{https://youtu.be/O8vevgfxxwQ}.

The virtual factory presents a foundational framework for exploring the evolution of manufacturing paradigms through immersive, interactive environments in extended reality. Building on its current capabilities, future work may focus on enhancing both the technological depth and pedagogical breadth of the platform. One promising avenue is the integration of advanced sensor-based AI to enable adaptive learning experiences and real-time feedback \cite{kim2024behavioral}. By analyzing user performance, communication patterns, and physiological responses, the system can deliver personalized guidance, identify bottlenecks in collaboration, and recommend task reallocations to improve team performance and efficiency.

There is also significant potential to extend the scope of the virtual factory beyond automotive production. Future modules could simulate production in aerospace, electronics, or biomedical sectors, each with domain-specific workflows, standards, and customization demands. This would provide broader applicability and increase the utility of the platform for a wider range of academic, industrial, and workforce development audiences. In addition, there are research opportunities to study team dynamics, cognitive load, and decision-making under varying complexity levels across paradigms. The integration of multimodal sensing, such as heart rate variability, gaze tracking, and galvanic skin response, will facilitate empirical investigations into human factors and ergonomics within smart manufacturing environments. 

Lastly, as XR technologies continue to advance, it is imperative to develop the next generation of blended human-cyber-physical systems towards smart manufacturing metaverse. Real-world factory must be seamlessly integrated with VR, AR, MR implementations, augmenting head-mounted displays with sensors and actuators for real-time interactions between digital and physical twins. In the state of the art, Oliver also highlighted the relevance and timeliness of VR technology in the context of manufacturing engineering and product development \cite{10.1115/1.1740774}. Data integration is stressed to be a critical issue during the process of transforming CAD models into large-scale digital assets required by XR applications. Specifically, ``virtual engineering" is conducive to leverage new XR user interfaces (i.e., in the immersive environment) for optimizing the parameter design, geometric prototype, and product development. 

Bordegoni et al. further showcased the development of a multisensory VR application that focuses on training machine operators about how to use a machine, as well as use personal protective equipment (PPE) for safety measures \cite{10.1115/1.4053075}. In addition, Ipsita et al. presented the VRFromX as a system workflow to enable domain users to create digital contents (or assets) by selecting the region of interest (ROI) in the scanned point clouds to build the virtual models. An attractive feature of VRFromX is an enabling tool that requires minimal programming skills and experience \cite{10.1115/1.4062970}. Moreover, Li et al. described the delivery, update and management of AR contents for ubiquitous manufacturing. Real-time data are captured and updated from machines to increase interactions between AR and physical systems \cite{Liongnee}. Kadavasal and Oliver presented the fusion of VR models with sensors and image processing capabilities to create the multimodal tele-operation interface that provides real-time interactions with the virtual environment via on-board sensors in the vehicle \cite{kadavasaloliver}. After all, sensor integration is indispensable for real-time and remote tele-operations in the design and development of DTs in extended reality. 

\section{Cognitive Digital Twin} \label{sec:CDT}
\subsection{Physical twin and DT examples} \label{subsec:CDT1}
  
Advanced sensing ushers in large data streams to build a digital twin, and enables cyber-physical interactions through the 3D immersive environment \cite{yang2019internet}. As a fit-for-purpose digital representation, DT offers a higher level of flexibility to develop cognitive intelligence and provide simultaneous feedback to the factory \cite{lee2023digitala,lee2023digitalb}. As aforementioned, our team developed a VR-based DT of physical twins (i.e., 2 collaborative robots and two 3D printers), which allows users to tele-operate robots and 3D printers via the metaverse (also see Fig. \ref{fig:exampleDT}). Real-time sensor signals (i.e., temperature, acoustics) are collected from 3D printers and synchronized into the VR DT for data analytics. After printing is completed, robots pick the parts and assemble into the final product. Further expansion of DT capabilities in the virtual factory environment will allow real-time synchronization with physical manufacturing systems. This bidirectional connectivity would support hybrid learning and training scenarios where users operate physical and virtual assets in parallel, reinforcing digital literacy and system-level thinking. 

	\begin{figure*}[hbt]
		\centering
		\includegraphics[width=5.5 in]{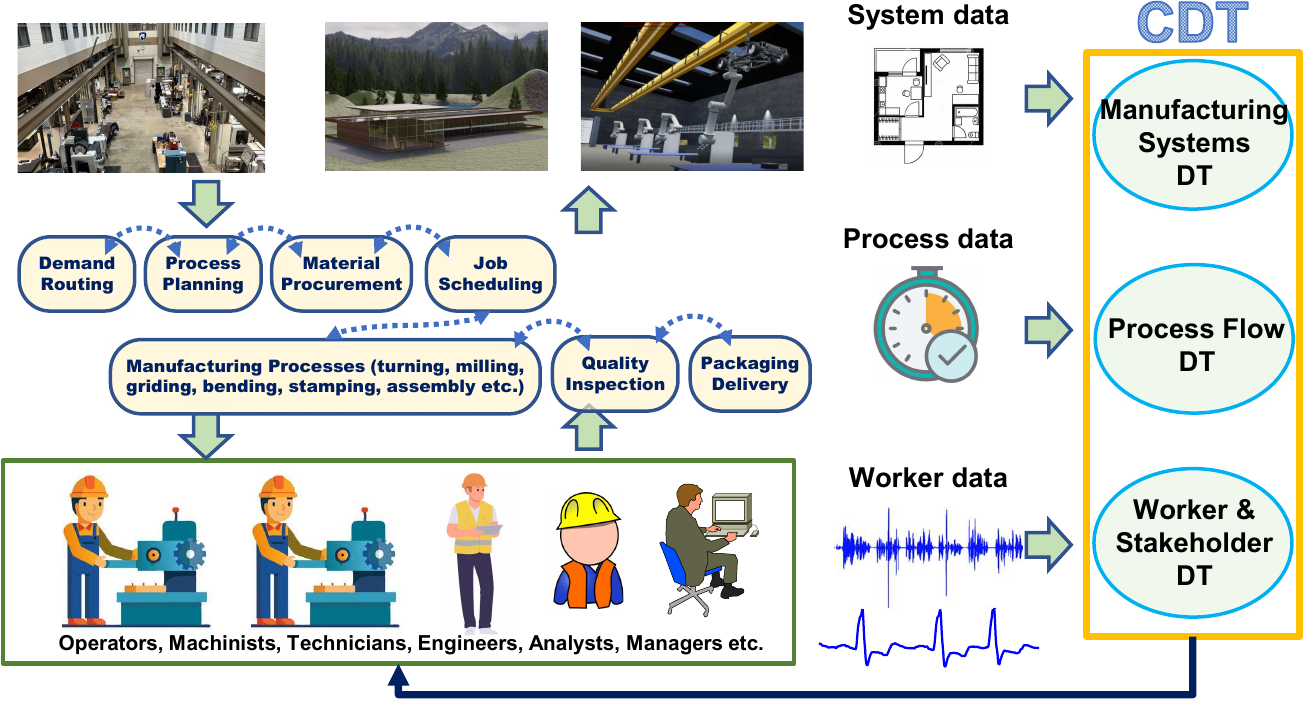}
		\caption{Illustration of CDT for worker training, process flow optimization, facility planning.}
		\label{fig:CDTworker}
	\end{figure*}

\subsection{Embedded AI for cognitive intelligence} \label{subsec:CDT2}
Manufacturing metaverse exists in the virtual world, which is persistent and self-evolving even if some physical twins are offline. Because a manufacturing system involves a variety of heterogeneous agents (e.g., human operators, users, jobs, machines, 3D printers, queues, and material handling robots), DTs of these agents need to develop self-learning capabilities to optimize the next-step decisions when physical twins are offline. Thus, \textbf{Cognitive Digital Twin} refers to \textit{the embedding of cognitive intelligence in a DT of a physical twin pertinent to a manufacturing agent} \cite{lee2025cognitive,lee2025cdt}.  In other words, CDT employs the agentic AI to enable multi-agent learning and decentralized optimization, thereby improving the smartness and autonomy of manufacturing metaverse. This is significant, because a hallmark of the manufacturing-as-a-service (MaaS) ecosystem is the distributed environment of users, assets, processes, and systems.    

Based on the theory of cognitive psychology \cite{solso2005cognitive}, CDT’s cognitive intelligence means that a digital agent is capable of sensory perception, situation awareness, evolutionary learning, decision optimization, and make reasoning to solve new problems in a different context. As a result, metaverse consists of a large network of agentic AIs that develop a high level of cognitive intelligence through the interaction process between digital and physical twins. For example, the CDT of a machine agent will perceive and learn from real-time data (e.g., operational statuses, queue length, and job plans) from the physical twin. In the case of machine offline, this agent can also run the simulations and perceive simulation model for decision support in the DT environment. Notably, as a part of machine networks, it will interact with other agents for multi-agent learning. With the embedded AI, each agent will develop self-autonomy and learn to optimize next-step decisions to maximize the cumulative reward function. However, it is not uncommon that these agents may have conflicting objectives. Therefore, they must also learn to cooperate, compete, negotiate and collaborate in the multi-agent ecosystem to balance individual rewards with the system-level key performance indicators (KPIs).   

\subsection{Human-centered CDT} \label{subsec:CDT3}
As shown in Fig. \ref{fig:CDTworker}, we envision a multi-level CDT that consists of digitalized human workers (and pertinent stakeholders), work process flows, as well as manufacturing systems, facilities. In other words, every step when processing a job, every unit in the facility, and every stakeholder, and real-time sensing observations and feedback are digitally represented and optimized.

• \textbf{Human DT}: Digital representations are constructed for key personnel involved in manufacturing processes such as operators, machinists, technicians, engineers, analysts, and managers. By simulating these stakeholders' interactions within realistic manufacturing scenarios, these digital models facilitate proactive assessment and improvement of work conditions. Specifically, they enable early identification of potential ergonomic risks, workload bottlenecks, cognitive overload, and safety hazards. The use of human DTs reduces mental workload, enhances physical safety, optimizes cognitive states, and improves overall job performance and productivity.

• \textbf{Process DT}: The process flow involves a series of steps such as demand routing, process planning, material procurement, job scheduling, job processing, quality inspection and package delivery. Process DT increases the visibility into the current state of manufacturing operations (e.g., work-in-process, lead time, energy consumption). This, in turn, help optimize process flows and guide workers to identify and solve bottlenecks beforehand, and further improve the KPIs.

• \textbf{Systems DT}: Human and process DTs are operating in physical and digital environments. In a manufacturing system, there are a large number of heterogeneous agents pertinent to assets and resources (e.g., machines, 3D printers, queues, and material handling robots). Agentic data and facility information, e.g., floor plans, asset layouts, building ambient environment, are required to be integrated into system DT and further develop AI and analytics to optimize the performance of human workers in the manufacturing system or a facility.

The seamless integration of multi-level CDT is conducive to realizing the future of human-centered MfgVerse and optimizing the manufacturing process flows. Hence, CDT design will require heterogeneous sets of data that are human related, process related, and facility/system related. Data analytics further help identify patterns, human behaviors, process deficiencies, and other variables that can be utilized in the CDT. More importantly, digital humans (or avatars) need to be built to represent the different roles in the manufacturing context. Our team has developed several avatars in Blender software and imported those into Unreal Engine. The avatars include detailed representations of the individual’s physical, cognitive behavior, as well as relevant environmental factors that may affect their work. Lastly, agentic AI needs to be developed and integrated into the CDT. The agentic models can also be used to continuously update and refine the CDT based on new data and observations.

	\begin{figure*}[hbt]
		\centering
		\includegraphics[width=6.5 in]{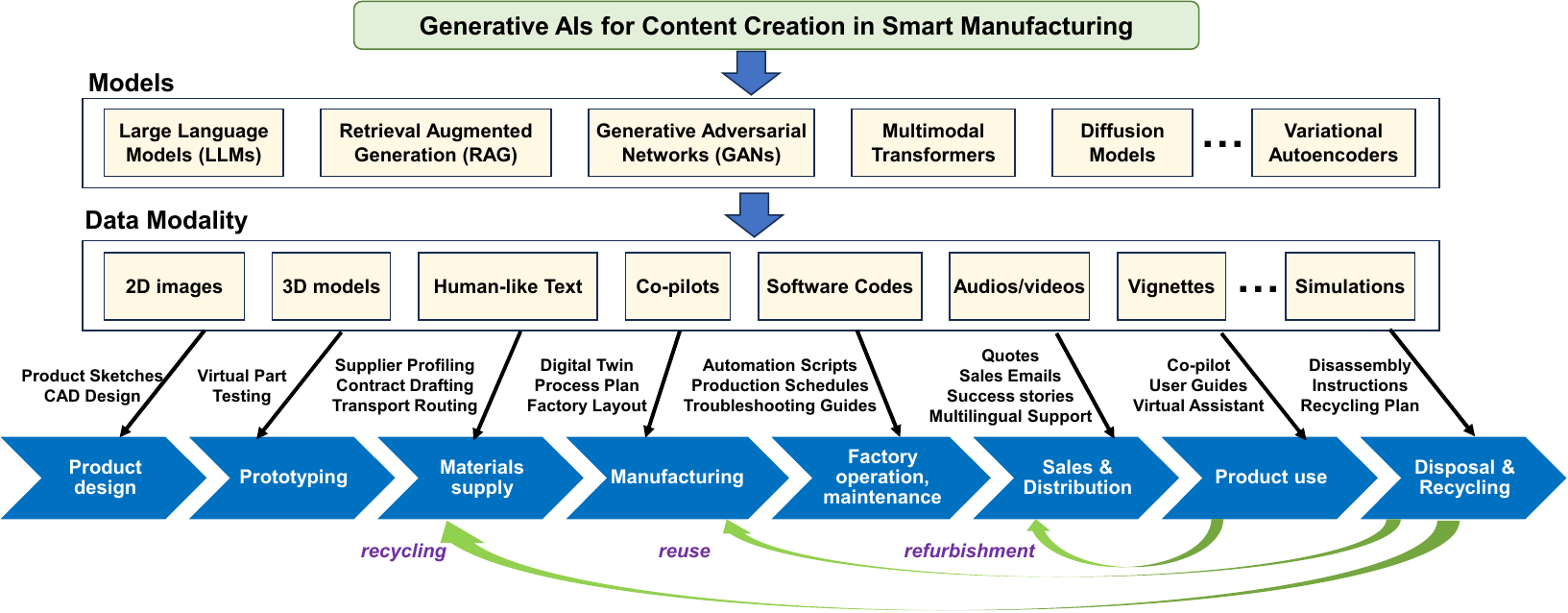}
		\caption{Generative AI for creative and automated generation of content and solutions for smart manufacturing.}
		\label{fig:genai}
	\end{figure*}

\textbf{Workforce Safety and Well-being}: Hence, human-centered CDT helps monitor and assess health conditions and fatigue levels of human workers in real time and identify changes in physical, cognitive, and emotional health toward human-centered MfgVerse. Notably, manufacturing metaverse provides a cyber-physical feedback loop where DT simulations, real-world sensor data, and AI insights jointly work together to enhance workforce safety and well-being toward creating a more human-centered workplace. For example, VR, AR and MR can create an immersive environment for safety and hazard training where workers practice the identification and response to simulated dangerous scenarios (e.g., chemical spills, fires, machine failures, or accidents) without risk. Also, workflows and safety procedures can be virtually tested without disrupting real production. Furthermore, wearable sensors can be used to help analyze worker movement to redesign ergonomic workflows that reduce repetitive strain or unsafe motions. In addition, the metaverse environment provides an unprecedented opportunity to perform AI-based sentiment analysis and leverage generative AIs (e.g., chatbots) and large language models (LLMs) to offer wellness check-ins and private mental health support. Metaverse technologies and AI play increasingly important roles to enhancing workforce safety and well-being in various ways from simulated emergency responses, wearable monitoring, ergonomic designs, wellness check-ins to mental health support.

\subsection{Generative AI for Smart Manufacturing} \label{subsec:GenAI}
Generative AI (GenAI) is also propelling the creative and automated generation of content and solutions to improve the smartness level of cyber-physical manufacturing systems. Note that GenAI embodies a broad range of models, learning algorithms, data modalities, and applications across the product life cycle in manufacturing \cite{li2025generative}. As the landscape is rapidly evolving, we cover only a sparse set of potential applications and implications pertaining to manufacturing. Fig. \ref{fig:genai} shows different types of models and learning algorithms behind GenAIs (e.g., large language models, retrieval-augmented generations, generative adversarial networks, multimodal transformers, diffusion models, and variational encoders). 

The choice of model architecture and learning algorithms is dependent, to a large degree, on the modality of data and application purposes. For example, large language models can be used as foundation models to receive further context-specific training for codes generation or serve as chatbots that provide 24/7 multilingual support for marketing and sales.  Retrieval-augmented generations are commonly employed to provide work instructions, maintenance guides, and troubleshooting steps. Generative adversarial networks play the two-player minimax game between generator and critic networks to generate realistic images, which can be used to create factory layouts or product sketches. In addition, transformers, diffusion models, and encoders are known to learn low-dimensional distribution of latent representations of 2D images or 3D designs. As such, they can be used for design creation, virtual prototyping, topology optimization, vignette generation, and design variant exploration.

However, GenAI needs extensive training on large amounts of data to learn the underlying distribution of complex-structured texts, codes, images, or 3D objects. It is often challenging to collect well-structured physical data from the factory and build such large databases due to practical and proprietary reasons. Fortunately, DT simulations provide a greater level of flexibility to generate large amounts of data for the GenAI training under different scenarios.  For example, DT-supported factory simulations can generate rich data about system configurations, factory layouts, and key performance indicators (KPIs) to train GenAI that can automate the decision tasks and configuration alternatives in response to unexpected disruptions or faults. After training, the speed and efficiency of GenAI are far exceeding the performance of traditional sequential design and simulation optimization. This is mainly due to the fact that GenAI can instantaneously provide human-like text through nature language interfaces to support the process of decision making. In contrast, traditional optimization approaches need to run computer simulations iteratively to search for the optimal solution during system disruptions or failures.

Indeed, GenAI is emerging as a game-changer to improve the smartness level of manufacturing. Nonetheless, there are also significant challenges and risks pertinent to the use of GenAI in manufacturing. For example, hallucination risks refer to the generation of false and misleading contents that are not grounded on facts or do not exist. Also, GenAI lacks interoperability with the factory manufacturing execution systems (MES) and enterprise resource planning (ERP) systems due to data security and proprietary protection. Hence, future research is necessary to integrate GenAI with manufacturing-specific knowledge base (e.g., graphs, databases, documents) or develop human-in-the-loop AI models to mitigate the hallucination risks. Lastly, there is an urgent need to develop a structured integration approach to increase interoperability by combining middleware bridges, system ontology, data semantics, user interfaces, and operational technology (OT) architecture.

\section{Future Directions, Challenges, and Research Opportunities} \label{sec:future}
Technological advances are rapidly transforming the landscape of future manufacturing. HMIs are changing from 2D monitor screens, smartphones, tablets to 3D immersive interactions via XR such as VR, AR, MR. Advanced sensing and control systems are streaming large amounts of data pertinent to workers, resources, assets, facilities, processes, systems to name a few for cyber-physical twinning and optimization. CDTs and agentic AIs improve the smartness and autonomous levels of manufacturing systems, ushering in a new paradigm of MaaS sharing economy. However, fully exploiting the emerging technologies of XR, cyber-physical twinning, distributed agentic AIs to develop smart and interconnected manufacturing metaverse depends, to a great extent, on addressing the following challenges.

• \textbf{Readiness of manufacturing workforce}: Manufacturing industries are facing a shortage of skilled workforce with digital literacy in XR, DT, AI and analytics. A survey indicated that manufacturing companies need around 3.8 million new employees in the U.S. by 2033, but experience difficulty filling approximately 1.9 million open positions ($\sim$50\% unfilled) \cite{deloitte2024}. Digital literacy is among the top list of three skill gaps according to the survey to manufacturers and employers. The future of manufacturing is highly dependent on STEM workforce who are skilled in digital technologies. There is an urgent need to design and develop relevant educational programs and curricular contents in U.S. institutions.

• \textbf{Metaverse cybersecurity and privacy}: MfgVerse involves an interconnected network of CDT agents (e.g., machines, robots, and operators) that communicate with each other, which also means an increasing level of vulnerabilities. Agentic AIs exploit large amounts of pertinent assets, processes, and systems data \cite{lee2024privacy}. As a result, cybersecurity and privacy concerns arise. There are different types of cyber attacks, e.g., privacy-leaking attacks, adversarial attacks, or model inversion attacks \cite{krall2020mosaic}. These attacks may bring significant disruptions to manufacturing operations, even endangering the intellectual property (e.g., facility layouts, process plans, job schedules) and causing ethical and legal concerns. There is an urgent need to develop privacy-preserving metaverse technologies (e.g., XR, DT, AI, cryptosystems) that make manufacturing smart and secure.

• \textbf{System interoperability, data synchronization, and real-time DT model calibration}: Manufacturing industries often have a mix of legacy and modern machines, different XR platforms, DT simulation models, and control systems. Standard protocols are lacked for the integration of heterogeneous agents in physical twins with digital agents in the 3D immersive environment. Also, sensors are commonly set with different sampling rates because there are slowly-varying and highly varying frequencies in the nature of different systems \cite{chen2015sparse}. Sensor failures and data latency disrupt the cyber-physical interactions, which can impact the fidelity of DT simulations \cite{meyers2021markov}. Lastly, multi-level DT models are often high-dimensional and computationally expensive, thereby posing significant challenges on continuous calibration and synchronization of DT models with real-time data streams.\\

In addition, managerial teams are more concerned about implementation costs, return on investment (ROI), new value propositions, new business models, and the shortage of skilled manpower for the manufacturing metaverse. No doubt, the success of MfgVerse is greatly dependent on helping manufacturers gain competitive advantages in the global market. These advantages include high monetary ROI, cost reduction, performance improvement, sustainable economy, and tangible values (e.g., XR learning factory for workforce training, remote work and collaboration, on-demand manufacturing services). Addressing these challenges will lead to new avenues of fundamental and applied research in the design of metaverse technologies.\\

\textbf{Opportunity 1. Learning factory for workforce training and safety in the metaverse}\\
The learning factory leverages the 3D immersive metaverse environment for training and education in advanced manufacturing, leveraging XR training modules and sensor technologies. See an example of VR learning factory in this YouTube link: \url{https://youtu.be/i5nVBOFaeWw}. Virtual learning factory offers a risk-free environment for remote training and continuous upskilling. In other words, new employees do not have to enter a physical factory but can be trained remotely for faster onboarding. Whenever new manufacturing technologies (e.g., laser powder bed fusion) are in place, experienced technicians can guide new employees on operational procedures in the virtual metaverse. In addition, "what-if" scenarios can be simulated to test how new employees respond and solve the problems, identify root causes, and find creative solutions.

\textbf{Opportunity 2. Distributed cryptosystems for MaaS sharing economy}\\
Networked and interconnected systems bring an increasing level of vulnerability. Manufacturers often decline the use of AI computing in sensitive data spread across different entities. There is a great opportunity to develop the privacy-preserving framework that will enable federated learning on siloed and encrypted data for smart manufacturing. It is critical to ensure that sensitive operational data remains protected throughout the analytical processes, promoting trust and security in the manufacturing metaverse \cite{kuo2024federated,wang2023security}. Note that distributed cryptosystem leverages the operations (e.g., homomorphic encryption, decryption, key generation, digital signature, or verification) to perform multi-agent distributed learning and computation, instead of aggregating the data to a single centralized authority \cite{krall2024distributed}. This offers an unparalleled advantage for research studies to realize distributed MaaS in the metaverse. In addition, blockchain technology enables decentralized banking, secure data certification, and online transactions for two-sided markets in the manufacturing sharing economy.

\textbf{Opportunity 3. Human factors and usability considerations in the manufacturing metaverse}\\
Human-centered smart manufacturing focuses on designing systems that enhance human capabilities and ensure worker well-being. Traditional manufacturing approaches often overlook human factors, leading to inefficiencies and safety concerns. The integration of advanced technologies addresses these limitations by improving productivity and ergonomics \cite{begum2025exploring}. The implementation of XR factory impacts human factors by influencing cognitive load, adaptability, and safety. Assessing usability through appropriate metrics and addressing ethical and accessibility concerns are essential for successful adoption. There are great opportunities in balancing technology integration with human-centered design, strategies for reducing cognitive overload, and ensuring inclusivity in manufacturing environments.

\textbf{Opportunity 4. High-performance computing for distributed agentic AIs and real-time calibration of DT models in manufacturing metaverse}\\
A large-scale network of heterogeneous agents co-exist and evolve in the manufacturing metaverse. Advanced sensing increases the information visibility about real-time statuses of “manufacturing agents” (e.g., parts, operators, machines, and material handling systems). For example, in additive manufacturing processes, coaxial melt-pool monitoring provides a unique opportunity to control the quality of AM build \cite{zhang2024engineering,liu2023multimodal}. A client can track and visualize real-time statuses of his parts via 3D immersive dashboards \cite{yang2022spatiotemporal}. Advanced sensing will open avenues of opportunity to synchronize and calibrate DT models with real-time data and further help predict the remaining useful life (RUL) of machines and optimize the maintenance schedules. These DT agents can learn from interactions between physical and virtual agents, and further optimize their strategies to achieve optimal decisions. The agentic AI approach effectively addresses the high-level complexity in the metaverse and improves the performance of manufacturing services \cite{lee2025mgai}. Nevertheless, the training and learning of agentic AIs is computationally expensive. This, in turn, creates an opportunity to leverage high-performance computing \cite{kan2018parallel} to speed up the data transformation, information processing, and intelligence distillation for smart manufacturing. 

\textbf{Opportunity 5. Sustainable manufacturing metaverse for circular economy}\\
Metaverse technologies (e.g., XR, CDT, AI) show strong promise to promote sustainable development and circular economy. For example, XR helps green design and virtual prototyping. In other words, eco-friendly design ideas can be prototyped and tested virtually through remote teaming and collaboration. The client can also track carbon footprints and visualize environmental impacts at every stage of production lines via the 3D immersive dashboards. CDT and AI can analyze the value stream mapping of energy consumption \cite{wang2020sensor}, identify bottlenecks, and optimize the sustainability in production lines. There are also great opportunities to design an online marketplace in the metaverse to support reuse, recycling, remanufacturing, and reduce waste through the e-transaction of digital and/or physical assets. 

\textbf{Opportunity 6. Manufacturing metaverse for customer relationship management (CRM)}\\
An attractive feature of manufacturing metaverse is the 3D immersive world that interacts with the real world. CDT collects and synchronizes the real-time data to calibrate models in the virtual world. This provides an opportunity to develop new business strategies to manage relationships with customers. For example, a client can not only evaluate the product design in a 3D immersive environment, but also join the co-design or reconfiguration of products. Manufacturers can also engage customers in virtual demonstrations of production facilities with 3D immersive headsets, instead of in-person visits.

\section{Discussion and Conclusions} \label{sec:conclusions}
This paper highlights the transformative potential of smart MfgVerse through the development of emerging technologies such as XR, DT, AI in the manufacturing industry. The rise of persistent metaverse environments (e.g., Microsoft Mesh, Meta Horizon, Roblox) shows how sustained engagement is possible -- a model that industrial XR platforms are beginning to adopt. We envision that this new MfgVerse co-exists in the virtual and real world, leveraging advanced sensing, cyber-physical twinning, and agentic AIs to learn and self-evolve even if human users are offline. There is an unprecedented opportunity to develop a next-generation MfgVerse paradigm as a new form of manufacturing practices in the future. 

Therefore, this paper presents a forward-looking perspective on the development of metaverse technologies and current efforts in the manufacturing domain. Future directions, challenges, and opportunities are also discussed. By addressing current challenges and leveraging emerging technologies, the manufacturing industry can achieve more efficient, safe, and human-centered production environments. The new MfgVerse framework is positioned as a strategic response that consists of several innovations as follows: (i) XR learning factories that deliver persistent, scalable, immersive workforce upskilling; (ii) Advanced sensing, cyber-physical twinning and agentic AIs that boost the performance improvement; (iii) MaaS platforms that unlock the new sharing economy of distributed, on-demand production without heavy capital investment. Indeed, the emergence of new HMIs transforms the way people interact with machines, internet, and computing devices into the 3D immersive virtual world with physical senses. We hope that this work can catalyze more in-depth and comprehensive studies relevant to manufacturing metaverse technologies. 

In particular, manufacturing metaverse is conducive to sustainable development by enabling the green transformation of the industry.  Manufacturing is known to account for a large carbon footprint in the economy. Metaverse technologies can alleviate environmental impact by reducing waste, enhancing energy efficiency, supporting remote operations, and empowering the workforce. For instance, virtual design and prototyping help reduce the material waste by testing the product and process before physical build. Tele-operations and remote collaboration reduce the need for in-person meetings, thereby cutting travel-related carbon emissions. AR can also be used to overlay digital assembly instructions or maintenance guidelines on top of physical machines, which reduces human errors, minimizes the rework and scrap materials, and prolongs the lifespan of machines. Additionally, human workers can inspect the product life cycles and recycling processes through XR simulations, supporting the reuse and end-of-life sustainability goals.

Metaverse technologies (including VR, AR, MR, and DTs) are rapidly advancing in the manufacturing industry. However, technology readiness levels (TRL) can vary in different use cases. Some use cases are already in the TRL level 8-9 (i.e., technology qualified and proven), while others are still emerging. For example, PTC offers the off-the-shelf Vuforia solutions for AR-assisted repair guidance, quality inspection, and workforce training\footnote{\url{https://www.ptc.com/en/products/vuforia}}. As another example, Bosch also provides AR/VR platforms for immersive training and maintenance activities to improve operational efficiency and productivity\footnote{\url{https://www.bosch-softwaretechnologies.com/en/services/enterprise-services/manufacturing-excellence/arvr/}}. However, DT and AI workflows are still in the developmental and experimental phase for most factories, even though they are growing rapidly due to the increasing availability of computing power and data infrastructure. Further, the integration of fully immersive metaverse factories with AI and DT is still emerging at the low TRL level as pilots and demos. In a nutshell, metaverse technologies are ready and scalable in some domains like training, monitoring, maintenance, and design, but full immersive manufacturing metaverse are still in its infancy to be developed and upscaled as the infrastructure matures. We call on industry leaders, researchers, and policymakers to pilot and refine this framework, turning MfgVerse from its infancy into critical infrastructures for future manufacturing. 

\begin{acknowledgment}
This work was supported in part by the National Science Foundation under the Grant No. IIS-2302833 (to FA on learning factory in extended reality), IIS-2302834 (to HY on learning factory in extended reality), CMMI-2531899 (to HY on genrative AI for smart manufacturing), and the Natural Sciences and Engineering Research Council of Canada (NSERC) Discovery Grant (to RZ on virtual reality). Any opinions, findings, or conclusions found in this paper are those of the authors and do not necessarily reflect the views of sponsors.
\end{acknowledgment}

\bibliographystyle{asmems4}
\bibliography{reflib_AM}

\end{document}